\documentclass[twocolumn,prX,nofootinbib]{revtex4-1}
\pdfoutput=1
\usepackage{amsmath,amssymb,amsfonts,eucal,mathrsfs,graphicx,tensor,color}
\usepackage{hyperref}
\DeclareMathAlphabet{\mathpzc}{OT1}{pzc}{m}{it}

\newcommand{\beqn}[1]{\begin{equation*}#1\end{equation*}}
\newcommand{\beql}[1]{\begin{equation}\label{#1}}
\newcommand{\eeq}{\end{equation}}
\newcommand{\eq}[1]{\eqref{#1}}

\newcommand{\lf}{\left}
\newcommand{\rf}{\right}

\newcommand{\ft}{\protect\footnote}

\newcommand{\rt}{\sqrt}
\newcommand{\fr}{\frac}
\newcommand{\tn}{\tensor}

\newcommand{\Ga}{\Gamma}
\newcommand{\ga}{\gamma}
\newcommand{\de}{\delta}
\newcommand{\ep}{\epsilon}
\newcommand{\tH}{\theta}
\newcommand{\ka}{\kappa}
\newcommand{\la}{\lambda}
\newcommand{\Sg}{\Sigma}
\newcommand{\om}{\omega}

\newcommand{\dl}{\partial}
\newcommand{\Dl}{\nabla}
\newcommand{\vecDl}{\vec{\nabla}}

\newcommand{\dr}{\mathrm{d}r}
\newcommand{\ds}{\mathrm{d}s}
\newcommand{\dt}{\mathrm{d}t}
\newcommand{\dOm}{\mathrm{d}\Omega}

\newcommand{\Gae}{G_{\ae}}
\newcommand{\GN}{G_\text{\sc n}}
\newcommand{\met}{\mathsf{g}}
\newcommand{\Rie}{\mathpzc{R}\,}
\newcommand{\ac}{\CMcal{S}}
\newcommand{\lag}{\mathscr{L}}
\newcommand{\ADM}{{\text{\sc adm}}}

\newcommand{\uhor}{\text{\sc uh}}
\newcommand{\cuh}{c_{\uhor}}
\newcommand{\cae}{c_{\ae}}
\newcommand{\ruh}{r_{\uhor}}
\newcommand{\auh}{(a\cdot s)_{\uhor}}
\newcommand{\Xuh}{|\chi|_{\uhor}}

\newcommand{\upbt}{\mathbf{t}}
\newcommand{\krpin}{k^{+}_{r\text{\sc(i)}}}
\newcommand{\krpout}{k^{+}_{r\text{\sc(o)}}}
\newcommand{\krmin}{k^{-}_{r\text{\sc(i)}}}
\newcommand{\krmout}{k^{-}_{r\text{\sc(o)}}}

\renewcommand{\Im}{\text{Im}}

\begin{document}
\title{Towards thermodynamics of universal horizons in Einstein-{\ae}ther theory}
\author{Per Berglund}\email{per.berglund@unh.edu}
\author{Jishnu Bhattacharyya}\email{jishnu.b@unh.edu}
\author{David Mattingly}\email{david.mattingly@unh.edu}
\affiliation{Department of Physics, University of New Hampshire, Durham, NH 03824, USA}

\begin{abstract}
Holography grew out of black hole thermodynamics, which relies on the causal structure and general covariance of general relativity. In Einstein-{\ae}ther theory, a generally covariant theory with a dynamical timelike unit vector, every solution breaks local Lorentz invariance, thereby grossly modifying the causal structure of gravity. However, there are still absolute causal boundaries, called ``universal horizons'', which are not Killing horizons yet obey a first law of black hole mechanics and must have an entropy if they do not violate a generalized second law. We couple a scalar field to the timelike vector and show via the tunneling approach that the universal horizon radiates as a blackbody at a fixed temperature, even if the scalar field equations also violate local Lorentz invariance. This suggests that the class of holographic theories may be much broader than currently assumed.
\end{abstract}

%\makeatletter
%\let\old@fpheader\@fpheader
%\renewcommand{\@fpheader}{\old@fpheader\hfill UNH-12-02}
%\makeatother
\maketitle
%***********************************************************************************************************************************************************************************************************
\section{Introduction}
How general is gravitational holography? Since holography is strongly tied to the behavior of horizons in general relativity, one may naively expect that holography is a feature of general relativity only, or at least extensions of general relativity that preserve the causal structure of the light cone. In this paper, we argue that this may not necessarily be the case -- one can formulate much of black hole thermodynamics, including a first law for and corresponding thermal radiation from at least spherically symmetric horizons, even for a theory that has no universal light cone and, in fact, violates the equivalence principle. Such a claim seems contradictory for a myriad of reasons; e.g., if one no longer has a universal light cone, even the notion of a well defined horizon is problematic. Also, without the equivalence principle, particles will not travel along paths defined by the geometry of the spacetime only, so identifying the temperature of a surface in spacetime with a geometric ``surface gravity'' seems impossible. However, as we show here, many of the existing concepts and constructions developed over the years for general relativity can be adapted for use in theories with modified casual structure.

Similar questions have been recently asked with two particularly relevant lines of inquiry. First, black holes have been extensively studied~\cite{HLBH} in Horava-Lifshitz gravity~\cite{HLG}, where the gravitational action is modified by the introduction of a preferred spacelike foliation, thereby breaking Lorentz invariance and significantly changing the causal structure of the theory. Without a corresponding modification of the matter action, the horizon of the black hole is the null Killing horizon, and one can investigate the behavior of this surface (cf.~\cite{Majhi:2009xh}), which indicates a violation of the entropy-area law familiar from general relativity. Such results are, however, inconsistent, as the existence of the preferred foliation in the gravitational sector will, via quantum corrections~\cite{Collins:2004bp, Iengo:2009ix}, generically violate Lorentz invariance in the matter sector as well. Therefore, the Killing horizon is no longer the causal horizon, and its entropy is not of interest.

This leads to the second line of reasoning, the nature of black hole thermodynamics in theories where Lorentz invariance is violated in the matter sector. In~\cite{Dubovsky:2006vk,Eling:2007qd}, the authors considered two matter fields with different speeds of propagation and showed that the Killing horizon splits into two horizons leading to perpetual motion machines and violations of the generalized second law. This is inconclusive, though, as generically there will be higher dimension Lorentz violating operators as well, and causal boundaries must be causal boundaries for all excitations, not just low energy ones, if you wish to apply a generalized second law. If one has causally modified gravitational and matter sectors in a theory and allows for higher dimension Lorentz violating operators in the matter sector, both pictures above change dramatically: there are causally inaccessible regions, the boundaries of such regions are the same for all fields but are not Killing horizons, and there exist first laws for these boundaries. In this paper, we show that the boundaries radiate thermally, thereby strengthening a possible thermodynamic interpretation, although open questions remain (which we address in the Discussions section below).

The gravitational theory we consider as our toy model is Einstein-{\ae}ther theory~\cite{JM:aetheory}, a generally covariant theory of the metric coupled to ``the {\ae}ther'', a unit timelike vector field $u^a$. The Lagrangian $\lag$ of the theory is given by
	\beql{lag:ae}
	\lag = \fr{1}{16\pi\Gae}(\Rie - \tn{Z}{^{ab}_{cd}}(\Dl_a u^c)(\Dl_b u^d) + \la(u^2 + 1))~,
	\eeq
with $\tn{Z}{^{ab}_{cd}} = c_1\met^{ab}\met_{cd} + c_2\tn{\de}{^a_c}\tn{\de}{^b_d} + c_3\tn{\de}{^a_d}\tn{\de}{^b_c} - c_4u^au^b\met_{cd}$. The $c_i$ are  arbitrary parameters of the theory and $\la$ is a Lagrange multiplier that enforces the unit constraint.

Einstein-{\ae}ther theory admits spherically symmetric ``black hole'' solutions in the following manner. Consider a static, spherically symmetric spacetime, and cover it with the  Painlev\'e (free-fall) coordinates such that the metric takes the form
	\beql{met:PGff}
	\ds^2 = -\dt^2 + \lf(\ga(r)\dt + f(r)\dr\rf)^2 + r^2\dOm_2^2~,
	\eeq
where $t$ is the Painlev\'e time function and $\ga(r) = \rt{1  + \chi_a\chi^a}$, $\chi^a \equiv \dl_t$ being the time translation Killing vector. Now let $\Sg_U$ denote a surface orthogonal to the {\ae}ther vector $u^a$, so that $U$ is the ``{\ae}ther time'' generated by $u^a$ that specifies each hypersurface in a foliation ($u^a$ is hypersurface orthogonal). Causality for matter fields is ensured by requiring that field excitations propagate to the future in $U$, so that no closed timelike curves are possible even though field excitations may travel superluminally. If one chooses $u^a$ such that at asymptotic spatial infinity $\chi^a$ and $u^a$ coincide, then as one moves in towards $r = 0$ each $\Sg_U$ hypersurface bends down to the infinite past in $t$, eventually asymptoting to a three-dimensional spacelike hypersurface on which $(u\cdot\chi) = 0$, which implies that the Killing vector $\chi^a$ becomes tangent to $\Sg_U$. This hypersurface is the universal horizon~\cite{Blas:2011ni}. It is a regular~\cite{BJS:aebh,EJ:aebh} causal boundary, as any signal must propagate to the future in $U$, which is necessarily towards decreasing $r$ at the universal horizon.

Since Einstein-{\ae}ther is generally covariant, one expects~\cite{Wald:NQ} the existence of a Smarr formula and corresponding first law of black hole mechanics. Such a law exists~\cite{BBM:mechuhor} for ranges of the $c_i$'s, $0 \leqq c_{14} < 2$, $c_{13} < 1$ and $2 + c_{13} + 3c_2 > 0$, where we use the notation $c_{14} = c_1 + c_4$, etc. These are sufficient constraints to ensure energetic stability and a good Newtonian limit as well. For two special choices of the coefficients, $c_{14} = 0$ and $c_{123} = 0$, analytic solutions have been found~\cite{BBM:mechuhor}, and for these solutions the first law takes the form
	\beql{exact-soln:1L}
	\de M_{\ae} = \fr{(1 - c_{13})\ka_{\uhor}\,\de A_{\uhor}}{8\pi\Gae}~,
	\eeq
where $M_{\ae}$, the total mass of the spacetime, is related to the ADM mass by $M_{\ae} = (1 - c_{14}/2)M_{\ADM}$, and $\ka_{\uhor}$ is the surface gravity at the universal horizon, i.e., $\ka = \rt{-\fr{1}{2}(\Dl_a \chi_b)(\Dl^a \chi^b)}$ evaluated at the universal horizon.

There is an additional benefit to these exact solutions. Spherically symmetric Einstein-{\ae}ther solutions possess an extra scalar {\ae}ther-metric degree of freedom~\cite{EJ:aebh}, which generically travels at a speed different from the speed of light~\cite{Jacobson:2004ts}. Outgoing matter radiation may therefore emit {\ae}ther-metric \v{C}erenkov radiation. For the exact solutions the speed of the {\ae}ther-metric mode goes to infinity ($c_{14} = 0$) or zero ($c_{123} = 0$). For infinite speed modes \v{C}erenkov radiation is forbidden, while for zero speed modes there is no energy lost~\cite{Frank:1937fk}, and so for these solutions \v{C}erenkov radiation can be ignored. The exact solutions also have the metric component $f(r) = 1$~\cite{BBM:mechuhor}.

A first law alone does not imply that universal horizons have a thermodynamic entropy proportional to the area associated with them. Since the universal horizon forms a causal boundary, one can imagine throwing objects through the universal horizon and argue that the generalized second law would be violated if the universal horizon had no entropy, similar to the standard argument in general relativity~\cite{Bekenstein:1973ur}. However, in order to concretely argue for a thermodynamic interpretation of the first law, one must at least show that the universal horizon radiates thermally.
%***********************************************************************************************************************************************************************************************************
\section{Radiation from the universal horizon}
In the tunneling approach for Hawking radiation from a stationary black hole, one considers particle pair creation near the event horizon~\cite{Damour:1976jd, PW:tunneling, Visser:tunneling}. The radiation is composed of positive energy outgoing particles (traveling forward in Killing time) that escape from just inside the horizon and negative energy ingoing particles (traveling backward in time) that fall into the black hole from just outside. Both these processes are forbidden classically, and therefore the quantum mechanical nature of the process is clear. A finite energy excitation measured at infinity is infinitely blueshifted near the event horizon and so the semiclassical limit (in the form of WKB or eikonal/Hamilton-Jacobi methods) is adequate for calculating the tunneling amplitude~\cite{PW:tunneling, Visser:tunneling}. In the following, we consider spherically symmetric radiation of a neutral scalar field using Painlev\'e-Gullstrand (PG) coordinates~\eq{met:PGff}, which are smooth everywhere for the exact solutions.

Let $\phi$ be a neutral scalar field governed by an action $\ac[\phi]$. In the semiclassical approximation, a given classical configuration $\phi(x)$ is interpreted as the wavefunction associated with the quantum state of a $\phi$-excitation, and is written as
	\beql{eikonal-ansatz:phi}
	\phi(x) = \phi_0\exp\lf\{i\ac[\phi(x)]\rf\}~,
	\eeq
where $\phi_0$ is a ``slowly varying" ($\equiv$ constant) profile, and $\ac[\phi(x)]$ is the scalar field action evaluated on the configuration $\phi(x)$. If $k_a$ is the four-momentum of such an excitation, then from the standard rules of quantum mechanics $-i\Dl_a\phi(x) = k_a\phi(x)$, whence one obtains the covariantized Hamilton-Jacobi equations
	\beql{eikonal-ansatz:HJE}
	k_a = \Dl_a\ac[\phi(x)]~.
	\eeq
Of course, \eq{eikonal-ansatz:HJE} does not have any dynamical content yet, because we still have not imposed any equation of motion. In the eikonal approximation, this is achieved by imposing an appropriate energy-momentum dispersion relation on $k_a$~\eq{eikonal-ansatz:HJE}; we will come back to this below.

Specializing to spherical symmetry, we make the standard ansatz for the phase of the field configuration~\eq{eikonal-ansatz:phi}
	\beql{eikonal-ansatz:phase}
	\ac[\phi(t, r)] = \mp\,\om t + \int^r\dr'k_r(r')~.
	\eeq
Comparing \eq{eikonal-ansatz:HJE} with \eq{eikonal-ansatz:phase} we see $(k\cdot\chi) = \mp\om$, i.e., $\om$ (which is positive by assumption here and henceforth) is the magnitude of the Killing energy of the excitation, and the top (bottom) signs refer to positive (negative) energy excitations, while $k_r(r)$ is the three-momentum of the excitation with respect to the free-fall observer.

The ansatz~\eq{eikonal-ansatz:phase} along with a dispersion relation allows us to solve for $k_r(r)$ in terms of $\om$ and the metric components. As we show below, the superluminal dispersion that we will consider has four physical solutions: $k^{\pm}_{r\text{\sc (i)}}(r)$ and $k^{\pm}_{r\text{\sc (o)}}(r)$, where $+(-)$ refers to positive (negative) energy and subscript {\sc i}({\sc o}) means in(out)going. By time reversal invariance we further have $\krpout(r) = -\krmin(r)$ and $\krpin(r) = -\krmout(r)$. Among these, $\krpout(r)$ and $\krmin(r)$ will be shown to be singular at the universal horizon (classically forbidden trajectories), while $\krpin(r)$ and $\krmout(r)$ will be smooth. The tunneling probability, given by $\Ga \sim \exp\lf[-2\Im\ac\rf]$, can then be evaluated by using~\eq{eikonal-ansatz:phase} as
	\beqn{
	2\Im\ac = \Im\lim_{\ep \to 0}\lf\{\int\limits_{\ruh - \ep}^{\ruh + \ep}\dr'\,\krpout(r') - \int\limits_{\ruh + \ep}^{\ruh - \ep}\dr'\,\krmin(r')\rf\}~,
	}
where $\ruh$ is the location of the universal horizon. The first term corresponds to the tunneling of a positive energy mode out of the black hole, while the second yields the corresponding negative energy tunneling in part. The imaginary parts of the integrals are due to the singularities on the contours of the integration. To evaluate the integrals, we push the contours below the singularity in the first integral and above the singularity in the second~\cite{PW:tunneling}. The imaginary part then effectively comes from the residue of a closed counterclockwise circuit encircling the singularity at the universal horizon
	\beql{eikonal-ansatz:2ImS}
	2\,\Im\ac = \Im\oint\dr\,\krpout(r)~.
	\eeq
If the right-hand side is linear in $\om$ (up to $\om$ independent chemical potential terms), then the emission is thermal.

We now need to specify the scalar field action in order to calculate the spectrum from the universal horizon. We wish to violate Lorentz invariance and examine higher dimension operators (while keeping the field equations second order in $U$-time derivatives), so we choose our model Lagrangian as
	\beql{lag:ae-phi:O4}
	\lag = -\fr{s_\phi^2}{2}\met_{(\phi)}^{a b}(\Dl_a\phi)(\Dl_b\phi) - \fr{(\vecDl^2\phi)^2}{2k_0^2}~,
	\eeq
where $\met_{(\phi)}^{a b} = \met^{a b} -  (s_\phi^{-2} - 1)u^a u^b$ and $\vecDl_a$ is the projected (spatial) covariant derivative on $\Sg_U$. The signs of the $s_\phi^2$ (squared low energy speed of the $\phi$-excitations) and $k_0^2$ terms are chosen so that all modes are propagating modes in flat space. This leads to the following dispersion relation in the {\ae}ther frame upon using~\eq{eikonal-ansatz:HJE} and~\eq{eikonal-ansatz:phase}
	\beql{DR:O4}
	k_u(r)^2 = \fr{k_s(r)^4}{k_0^2} + s_\phi^2k_s(r)^2 + \fr{[\Dl_sk_s(r) + \hat{k}k_s(r)]^2}{k_0^2}~,
	\eeq
where $-k_u(r) \equiv -(u\cdot k)$ and $k_s(r) \equiv (s\cdot k)$ are the {\ae}ther frame energy and momenta of the excitation, respectively, $s^a$ is the unit spacelike vector orthogonal to $u^a$ (and so is parallel to $\chi^a$ at the universal horizon), $\Dl_s \equiv s^a\Dl_a$, and finally $\hat{k}$ is the trace of the extrinsic curvature of the two-spheres of constant $r$ and $t$ embedded in $\Sg_U$~\cite{BBM:mechuhor}.  There are obviously a whole tower of operators that could be added to the Lagrangian~\eq{lag:ae-phi:O4} which yield different dispersion relations and satisfy our above requirements; we choose the lowest two operators for simplicity. Another important point is that all propagating matter excitations with positive (negative) Killing energy must have positive (negative) {\ae}ther frame energy everywhere as well, since by~\eq{DR:O4} the four-momentum would otherwise have to vanish somewhere which is unphysical for a propagating mode\ft{This can be seen as follows: Consider a positive Killing energy excitation. Its {\ae}ther frame energy equals its Killing energy at infinity. Now, if its {\ae}ther frame energy is negative somewhere in the bulk, then it needs to vanish somewhere before that. By~\eq{DR:O4} $k_s(r) = 0$ at that point, and hence the mode has a zero four-momentum which is unphysical for a propagating mode. The same argument applies to negative Killing energy excitations.}.

To solve~\eq{DR:O4} and evaluate~\eq{eikonal-ansatz:2ImS} eventually, we need to relate $k_r(r)$, $k_u(r)$, and $k_s(r)$. Using \eq{eikonal-ansatz:HJE} and \eq{eikonal-ansatz:phase}, we find
	\beql{reln:kr-ks}
	k_u(r) {=} \fr{\pm\om {+} k_s(r)(s\cdot\chi)}{(u\cdot\chi)}, \;\,
	k_r(r) {=} \fr{\pm\om\sinh\tH {+} k_s(r)}{(-u\cdot\chi)},
	\eeq
where $\tH \equiv \tH(r)$ is a position-dependent boost angle relating the four-vector $\upbt^a$ defining the free-fall observer to the {\ae}ther frame according to $\upbt^a = \cosh\tH u^a - \sinh\tH s^a$. At the universal horizon, $\sinh\tH_{\uhor} = \Xuh^{-1}$.

Since we only need to extract the residue of $k_r(r)$ for the appropriate in/outgoing mode at $r = \ruh$~\eq{eikonal-ansatz:2ImS}, a Laurent series solution of~\eq{DR:O4} around the universal horizon is sufficient. Now~\eq{DR:O4} is a fourth-order equation and generally has four solutions (all with positive Killing energy). As we discuss below, only two out of these solutions have positive {\ae}ther frame energy and are therefore physically meaningful; they will be identified as $\krpin(r)$ and $\krpout(r)$, respectively. Using the $U$-time reversal invariance of~\eq{DR:O4} we can then find the corresponding negative Killing energy solutions, $\krmout(r)$ and $\krmin(r)$, respectively, by switching $\om \to -\om$ and $k_s(r) \to -k_s(r)$.

For the positive energy ingoing mode, $k_s(r)$ must be regular at the universal horizon. This regularity requirement fed into~\eq{DR:O4} yields $k_s(\ruh) = -\om\Xuh^{-1}$, showing that we do have an ingoing mode. Also, as indicated above, there are two regular solutions with the same value of $k_s(\ruh)$ but with $k_u(\ruh)$ differing by a sign. We can then discard the solution with negative {\ae}ther energy (but positive Killing energy) as being unphysical, as it cannot represent a propagating solution everywhere in the bulk. Note, by~\eq{reln:kr-ks}, $[(u\cdot\chi)k_r(r)]_{\uhor} = 0$, showing that $k_r(r)$ is finite at the universal horizon for the regular modes.

We now turn to the remaining two solutions of~\eq{DR:O4} for which $k_s(r)$ must be singular at the universal horizon. This is captured by the ansatz
	\beql{ansatz:ks(r):sing}
	k_s(r) = \fr{b(r)}{(-u\cdot\chi)^m}, \qquad m > 0, \qquad b(\ruh) \neq 0~,
	\eeq
where $b(r)$ is some function that is finite at the universal horizon and $m$ is the largest positive real number such that $[(-u\cdot\chi)^mk_s(r)]_{\uhor}$ is finite. From~\eq{ansatz:ks(r):sing} one can now prove that the $k_s(r)^4$ piece is the most singular piece on the right-hand side of~\eq{DR:O4} near the universal horizon. Hence, there is an approximate scale invariance characterized by a Lifshitz exponent $z = 2$ for the scalar field near the universal horizon. Continuing the analysis further, we finally conclude that~\eq{DR:O4} is satisfied if and only if
	\beql{eikonal-ansatz:m-buh:O2z}
	m = 1, \qquad b(\ruh) = \pm k_0\Xuh~.
	\eeq
For the negative solution of $b(\ruh)$, the excitation has negative {\ae}ther frame energy near the universal horizon and hence is unphysical. Therefore we must restrict $b(r)$ to be strictly positive at (and outside) the universal horizon. In this manner~\eq{ansatz:ks(r):sing} and~\eq{eikonal-ansatz:m-buh:O2z} (with $b(\ruh)$ positive) correspond to the positive energy outgoing excitation; the singular nature of $k_s(r)$ is very much expected, as this is the mode that is tunneling out through the universal horizon. Finally, by plugging \eq{ansatz:ks(r):sing} into \eq{reln:kr-ks} and invoking time reversal, the physical solutions of~\eq{DR:O4} are
	\beql{eikonal-ansatz:krpin-krmout:O2z}
	\krpout(r) = -\krmin(r) = \fr{\om\sinh\tH}{(-u\cdot\chi)} + \fr{b(r)}{(-u\cdot\chi)^2}~.
	\eeq
Hence, we have identified all the physical solutions of~\eq{DR:O4}.

The solutions~\eq{eikonal-ansatz:krpin-krmout:O2z} contribute to $2\,\Im\ac$ in Eq.~\eq{eikonal-ansatz:2ImS}. We perform a Laurent expansion of~\eq{DR:O4} around $r = \ruh$, solve for $b'(\ruh)$, and apply Cauchy's integral formula to compute the residue, which depends on $b(\ruh)$ {\it and} $b'(\ruh)$. Putting everything together, we finally find
	\beql{eikonal-ansatz:2ImS:O4}
	2\,\Im\ac =
	\fr{\om}{T_{\uhor}} + \fr{2 \pi\cae k_0\ruh} {N}~,
	\eeq
where
	\beql{def:T_uhor}
	T_{\uhor} = \fr{\auh\Xuh}{4\pi} = \fr{\cae}{4\pi\ruh} = \fr{(\cae/\cuh)}{8\pi\GN M_{\ae}}~.
	\eeq
Here $\auh$ is the magnitude of the acceleration $\Dl_u u^a$ evaluated on the universal horizon, $\cuh = \fr{1}{2}, \; \fr{3}{4}$ and $N= 1, \; 3\rt{2}$ for the $c_{123} = 0$ and the $c_{14} = 0$ solutions, respectively, $\cae$ is given by
	\beql{def:cae}
	\cae = \fr{1}{2}\rt{\fr{1}{\cuh}\lf(\fr{2 - c_{14}}{1 - c_{13}}\rf)}~,
	\eeq
and, finally, $\GN$ is the Newton's constant, related to $\Gae$~\eq{lag:ae} by $\Gae = \GN(1 - \fr{1}{2}c_{14})$~\cite{Carroll:2004ai}. Since the solutions at hand depend on a single parameter (the mass), one can further write $T_{\uhor} = (4\pi\cae)^{-1}\ka_{\uhor}$, thereby making a contact with the first law~\eq{exact-soln:1L}. It is, however, unclear whether associating the temperature with the surface gravity is natural for a (non-Killing) universal horizon.

The tunneling probability is $\Ga \propto e^{-2\,\Im\ac}$; therefore, in terms of the chemical potential $\mu_0 = -\cae^2 k_0/2N$,~\eq{eikonal-ansatz:2ImS:O4} leads to $\Ga \sim e^{-(\omega - \mu_0)/T_{\uhor}}$. By detailed balance, this yields a thermal spectrum~\cite{PW:tunneling, Visser:tunneling}, with a temperature given by~\eq{def:T_uhor}.
%***********************************************************************************************************************************************************************************************************
\section{Discussions}
Previous studies of Lorentz violating black hole thermodynamics argued for a violation of the generalized second law with {\it only} different speeds for fields. Here we have included higher derivative Lorentz violating terms in the matter action~\eq{lag:ae-phi:O4} which changes the nature of the causal boundary appropriate for generalized second law arguments and how Lorentz violation affects the emission spectrum. The spectrum remains thermal, even if fields have different values of $k_0$ -- only the effective chemical potential $\mu_0$ changes for each field. This is possible, as for any $k_0$ the universal horizon remains the unique causal boundary for high frequency modes, so the spectrum is dictated by the nearby local geometry.

While the first law and this result are suggestive, there are still open questions. First, the issue of reprocessing near the Killing horizon~\cite{Brout:1995wp, Corley:1996ar, Unruh:2004zk} is very important, as previous work has shown that the WKB approximation for low frequency modes breaks down near the Killing horizon even in the presence of Lorentz violation. Indeed, one can examine numerically the validity of the WKB approximation for our modes propagating on our exact solutions and the approximation also breaks down at the Killing horizon as the Killing frequency $\om$ becomes less than $k_0$, which indicates that significant further processing of low frequency modes may occur there. It is therefore possible that an observer at infinity would see a split or otherwise modified spectrum, which is qualitatively similar to recent results obtained by Busch and Parentani~\cite{Busch:2012ne} for Lorentz violating fields with de Sitter horizons. However, the further processing of low frequency modes by the Killing horizon is effectively a graybody factor and does not necessarily modify the essential nature of the universal horizon thermodynamics. Second, it is also possible that the universal horizon and the interplay of the thermal emission and Killing horizon reprocessing does not save one from the generalized second law violation arguments presented for two-speed Lorentz violating theories, which may indicate an instability of the universal horizon~\cite{Blas:2011ni}. We will return to these issues in future work.

\begin{acknowledgments}
We thank Sayandeb Basu, Ted Jacobson, Diego Blas, and Sergey Sibiryakov for useful discussions and feedback. As well, we especially thank Stefano Liberati and Renaud Parentani for reading drafts of this paper and providing helpful commentary. The work of P.B. is supported by NSF Grants No. PHY-0645686 and No. PHY-1207895. D.M. thanks the University of New Hampshire for research support. J.B. thanks the University of New Hampshire for support through a Dissertation Year Fellowship.
\end{acknowledgments}
%**********************************************************************************************************************************************************************************************************

\end{document}